\documentclass[
aps,prd,
preprintnumbers,
amsmath,
amssymb,
nofootinbib,
preprintnumbers
]{revtex4-2}
\usepackage{footnote}
\usepackage{braket}
\usepackage{amsmath}
\usepackage{graphics}
\usepackage{amsfonts}
\usepackage{amssymb}
\usepackage{latexsym}
\usepackage{color}
\usepackage{graphicx}
\usepackage{hyperref}
\usepackage{array,multirow}
\usepackage{physics}

\begin{document}
\title{Remote Hawking-Moss instanton and the Lorentzian path integral}
\author{Daiki Saito$^{1}$}
\author{Naritaka Oshita$^{2,3,4}$}
\affiliation{$^{1}$Division of Science, Graduate School of Science, Nagoya
University, Nagoya 464-8602, Japan}

\affiliation{$^{2}$Center for Gravitational Physics and Quantum Information, Yukawa Institute for Theoretical Physics,
Kyoto University, Kitashirakawa Oiwakecho, Sakyo-ku, Kyoto 606-8502, Japan}
\affiliation{$^{3}$The Hakubi Center for Advanced Research, Kyoto University,
Yoshida Ushinomiyacho, Sakyo-ku, Kyoto 606-8501, Japan}
\affiliation{$^{4}$RIKEN iTHEMS, Wako, Saitama, 351-0198, Japan}
\preprint{YITP-24-108}
\preprint{RIKEN-iTHEMS-Report-24}

\begin{abstract}
The Hawking-Moss (HM) bounce solution implies that the tunneling amplitude between vacua is uniquely determined by the vacuum energy at the initial vacuum and the top of a potential barrier, regardless of the field distance between them $\Delta \phi$. This implausible conclusion was carefully discussed in [E. J. Weinberg, Phys. Rev. Lett. 98, 251303, (2007)], and it was concluded that the conventional HM amplitude is not reliable for a transition to the top of distant local maxima (hereinafter referred to as the remote HM transition).
We revisit this issue and study the impact of the quantum tunneling effect on the remote HM transition. 
We demonstrate that the amplitude for such a distant transition is indeed smaller than the conventional HM amplitude by employing the Lorentzian path integral in a simple setup.
We consider a linear potential, which allows for analytic treatments, and evaluate the up-tunneling probability of a homogeneous scalar field in de Sitter spacetime.
The Picard-Lefschetz theory is employed to identify the relevant Lefschetz thimble, representing the relevant tunneling trajectory.
We then compare the resulting transition amplitude with the conventional HM amplitude.
We find that when the field separation $|\Delta \phi|$ is larger, the quantum-tunneling amplitude, estimated by our Lorentzian path integral, is smaller than that of the conventional HM amplitude.
This implies that the transition amplitude may be significantly suppressed if the thermal interpretation is not applicable and the quantum-tunneling effect is dominant for the remote HM transition.
\end{abstract}

\maketitle

\section{Introduction}
\label{Intro}

Vacuum phase transition is one of the most important phenomena in cosmology. The decay of a false vacuum is a quantum-tunneling process of a field $\phi$ from a metastable false vacuum to a stable true vacuum in the effective potential $V(\phi)$ at lower or zero temperatures.
This type of phase transition was first discussed in Refs.~\cite{Coleman:1977py, Callan:1977pt}, where the transition rate was evaluated by computing the on-shell Euclidean action, which corresponds to the non-trivial configuration of the field in the Wick-rotated space.
Most of the subsequent studies have been conventionally following the Euclidean path integral proposed in Refs.~\cite{Coleman:1977py, Callan:1977pt}.
One of the simplest models of the transition in curved spacetime derived using the Euclidean path integral method is the Coleman-De Luccia (CDL) bounce~\cite{Coleman:1980aw}, which respects the $O(4)$ symmetry.
When a positive large cosmological constant governs the background and a potential barrier is shallower, stochastic fluctuations can be dominant in the transition process. Indeed, in de Sitter (dS) background, the quantum fluctuations of long wavelength modes in a dS patch are squeezed and may be treated as stochastic fluctuations~\cite{Polarski:1995jg,Kiefer:1998qe,Kiefer:1998jk,Lyth:2006qz}, which is often referred to as "the quantum-to-classical transition" of fluctuation. 
The most famous instanton describing such a transition is known as the Hawking-Moss (HM) instanton \cite{Hawking:1981fz,Vilenkin:1983xq}. The transition amplitude of the HM bounce can be indeed reproduced by solving the Fokker-Planck equation in an equilibrium condition \cite{Linde:1991sk}. Such a process is often interpreted as a thermal hopping process~\cite{Weinberg:2006pc,Brown:2007sd,Oshita:2016oqn,Miyachi:2023fss}.

\begin{figure}[t]
\centering
\includegraphics[width=0.5\linewidth]{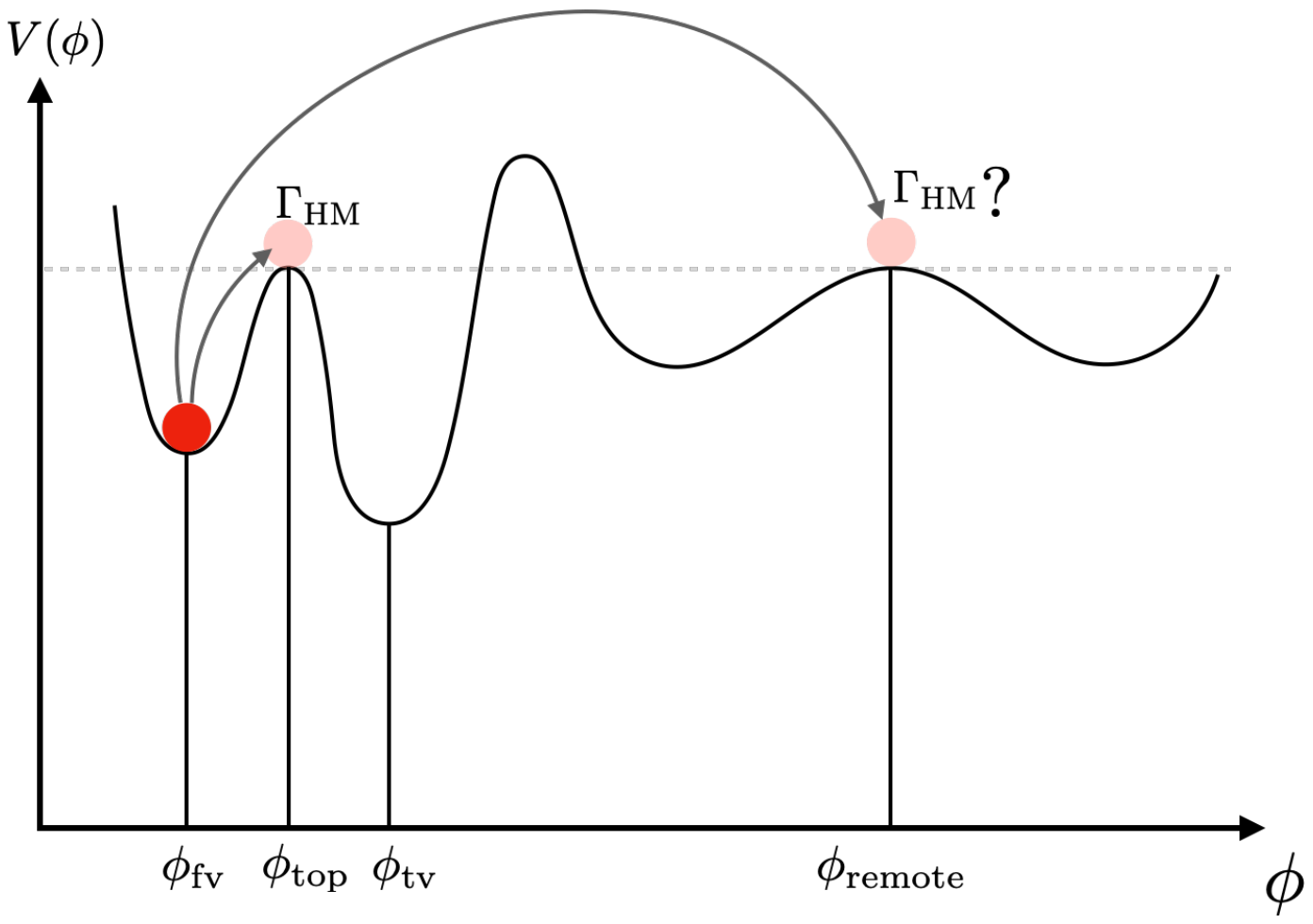}
\caption{A schematic picture showing distinct HM transitions, $\phi_{\rm fv} \to \phi_{\rm top}$ and $\phi_{\rm fv} \to \phi_{\rm remote}$. The two final states have the same potential height and hence give the same conventional HM transition amplitude. The main issue we discuss is whether the two indeed occur with the same transition amplitudes or not, which was first discussed in Ref.~\cite{Weinberg:2006pc}.}
\label{pic_schematic}
\end{figure}

The CDL bounce exists only when the potential barrier satisfies the following condition at the top of the barrier~\cite{Balek:2004sd,Hackworth:2004xb,Weinberg:2005af}
\begin{equation}
|V''(\phi)| \gtrsim \frac{1}{4}\left( \frac{8 \pi V}{3} \right),
\end{equation}
where the prime denotes the derivative with respect to $\phi$. We take the natural unit of $G = c= 1$ throughout the paper.
On the other hand, if the background is dS and there are local minima and maxima in $V(\phi)$, the HM bounce may {\it always} exist and are populated with the conventional HM probability as long as we rely only on the Euclidean path integral.
However, this interpretation is questionable and leaves the following two issues (for relevant discussions, see, e.g. Refs.~\cite{Linde:1991sk,Weinberg:2006pc}).
(I) The HM transition amplitude is given by
\begin{equation}
\Gamma_{\rm HM} = \exp \left[ \frac{1}{\hbar}\frac{3}{8} \left( \frac{1}{V(\phi_{\rm top})} - \frac{1}{V(\phi_{\rm fv})} \right) \right]
\label{conventional_HM}
\end{equation}
and does not depend on the field interval between a local minimum and maximum in the potential. Is the HM amplitude reliable even in the transition to the top of distant local maxima (see Fig.~\ref{pic_schematic})? Hereinafter such a transition is referred to as the remote HM transition in accordance with Ref.~\cite{Brown:2007sd}.
(II) Considering the HM transition as a stochastic process in the potential, the condition $|V''| \ll H^2$ should be imposed to derive the HM transition amplitude from the Fokker-Planck equation, where $H$ is the Hubble parameter. However, this condition is not necessary in the Euclidean approach~\cite{Linde:1991sk}.
Can we consider the HM transition even for $|V''| \gtrsim H^2$? If this is the case, how small is the corresponding transition amplitude?

In Ref.~\cite{Weinberg:2006pc}, it was argued that the HM amplitude can be understood as the Boltzmann factor, i.e., a probability of finding the homogeneous configuration at the top of maxima in the associated temperature~\cite{Weinberg:2006pc, Brown:2007sd, Oshita:2016oqn}.
This interpretation should be valid only after the whole configuration is in a thermal equilibrium state. For the remote HM transition, it is not clear which thermal or quantum-tunneling effect is more important.
As a non-equilibrium state, which could be involved in the case (I) as we will discuss later, is hard to describe in general, we here consider the contribution of the quantum tunneling effect to the remote HM transition. Also, the thermal effect would be suppressed when we address the issue (II).
We will then study issues (I) and (II) by considering only the tunneling effect due to the long-wavelength quantum fluctuation while turning off the thermal/stochastic noises.

In this paper, we discuss the two issues raised above by utilizing the Lorentzian path integral with the Picard-Lefschetz theory.
Recently, a method to evaluate the transition amplitude in the Lorentzian path integral was proposed in the context of quantum cosmology~\cite{Feldbrugge:2017kzv}.
In Ref.~\cite{Feldbrugge:2017kzv}, the authors applied the Picard-Lefschetz theory to find the relevant saddle point in the action, which is important to make the integration convergent.
This important technique to perform the Lorentzian path integral was also applied to the radio astronomy \cite{Feldbrugge:2019fjs}, the tunneling problem in quantum mechanics~\cite{Tanizaki:2014xba,Cherman:2014sba,Ai:2019fri,Matsui:2021oio,Feldbrugge:2023frq}, the thin-shell vacuum decay process~\cite{Hayashi:2021kro} and so on (\cite{DiazDorronsoro:2017hti,Baldazzi:2019kim,Rajeev:2021zae}).
We use this technique to see the quantum tunneling effect in the up-tunneling process.
One may think that the quantum tunneling effect on the homogeneous transition is negligible as the primordial fluctuation may be decohered on the superhorizon scale.
However, the decoherence effect may highly depend on how we split the whole system into an observed system and environment, which involves some artificial choice, more or less. In this sense, the quantum nature of the fluctuation may not vanish even in the superhorizon, and it could affect the subhorizon dynamics. The detailed discussion on this is beyond our scope in this paper.

This paper is organized as follows.
In Section~\ref{Euclid}, we briefly review the Euclidean path integral formalism for the tunneling problem.
The method to compute the transition rate using the Lorentzian path integral is discussed in Section~\ref{Lorentzian}.
There, we particularly focus on spatially homogeneous tunneling in a dS spacetime.
The calculation of transition probabilities using a specific example of effective potential will be carried out in Section~\ref{Linear}.
Finally, we present our discussion and conclusions in Section~\ref{Discuss}.

\section{Euclidean formalism: CDL bounce and HM bounce}
\label{Euclid}

In this section, we review the Euclidean path integral to evaluate the transition probability of a homogeneous scalar field in the semi-classical regime. Throughout this section, we assume that the geometry has the $O(3,1)$ symmetry and the spacetime metric is given by
\begin{align}
    ds^2=-dt^2+\rho^2(t)d\Omega_{3}^2, \label{eq:metric}
    \end{align}
where $d\Omega_{3}$ is the line element of the unit three-sphere. We shall consider the canonical real scalar field $\phi$ with the action of
\begin{align}
    S[\phi]&= \int d^4x\sqrt{-g}\qty[\frac{R}{16\pi }-\frac{1}{2}\partial_{\mu}\phi\partial^{\mu}\phi-V(\phi)]+S_{B},
    \end{align}
where $V(\phi)$ and $S_{B}$ are the scalar potential and the Gibbons-Hawking-York term, respectively. 

We here review the evaluation of the tunneling probability in the Euclidean path integral method~\cite{Coleman:1977py,Coleman:1980aw,Hawking:1981fz}.
In order to apply the Euclidean path integral method, we shall perform the Wick rotation: $t\rightarrow\xi:=it$.
Then, the Euclidean action can be written as
\begin{align}
   S_{E}[\phi]&=2\pi^2\int d\xi\qty[\rho^3\qty{\frac{1}{2}\qty(\frac{d\phi}{d\xi})^2+V(\phi)}-\frac{3}{8\pi }\qty{\rho+\rho\qty(\frac{d\rho}{d\xi})^2}].
   \label{eq:SE}
\end{align}
Varying the action with respect to $\phi$ and $\rho$, we obtain the equations of motion (EoM) of the scalar field and the Hamiltonian constraint as
\begin{align}
   &\frac{d^2\phi}{d\xi^2}+\frac{3}{\rho}\frac{d\rho}{d\xi}\frac{d\phi}{d\xi}=\frac{dV}{d\phi},
   \label{eq:phieq}\\
   &\qty(\frac{d\rho}{d\xi})^2=1+\frac{8\pi}{3}\rho^2\qty{\frac{1}{2}\qty(\frac{d\phi}{d\xi})^2-V}.  
   \label{eq:rhoeq}
\end{align}
Since the system depends only on the Euclidean time $\xi$, Eq.~\eqref{eq:rhoeq} is the only nonvanishing component of the Einstein equations.
According to the Euclidean path integral, the vacuum decay rate, $\Gamma$, can be computed as
\begin{align}
\Gamma&\propto e^{-\mathcal{B}/\hbar} \\
\mathcal{B}&:=S_{E}\qty[\bar{\phi}_{E}]-S_{E}\qty[\phi_{\textmd{FV}}],
\end{align}
where $\phi_{\textmd{FV}}$ is the field value of the false vacuum.
In Ref.~\cite{Coleman:1980aw}, the transition from false vacuum to true vacuum in the spacetime~\eqref{eq:metric}, i.e., the CDL bounce solution, was discussed.

The nucleation of the CDL bubble might be the most preferred transition process in the homogeneous cosmological background. 
However, the CDL solution does not exist if the scalar potential satisfies
\begin{align}
\left|\frac{d^2V(\phi_{\textmd{top}})}{d\phi^2}\right|<\frac{32\pi V(\phi_{\textmd{top}})}{3}, 
\label{eq:Vcond}
\end{align}
where $\phi_{\textmd{top}}$ denotes the field value at the top of the potential barrier. 

Even when the condition~\eqref{eq:Vcond} is satisfied, and there is no CDL bounce, we can always obtain the homogeneous solution called the HM bounce~\cite{Hawking:1981fz}.
The solution corresponds to a static scalar field configuration,
\begin{align}
\frac{d\phi}{d\xi}=0=\frac{d^2\phi}{d\xi^2},
\label{time_derivative}
\end{align}
which means that the field sits on the potential top $\left.dV/d\phi\right|_{\phi_{\textmd{top}}}=0$.
Then, the Einstein equation can be rewritten as
\begin{align}
  \qty(\frac{d\rho}{d\xi})^2=1-\frac{8\pi}{3}\rho^2V, 
\end{align}
and can be solved as
\begin{align}
  &\rho=\frac{1}{H_{s}}\sin(H_{s}\xi), \label{eq:rhoHM} \\
  &H_{s}:=\sqrt{\frac{8\pi V(\phi_{\textmd{top}})}{3}}. \label{eq:Hs} 
\end{align}
The solution~\eqref{eq:rhoHM} is the scale factor for the Euclidean dS spacetime.
By substituting it into Eq.~\eqref{eq:SE}, we obtain
\begin{align}
   S_{E}\qty[\phi_{\textmd{top}}]&=-\frac{3}{8\pi V(\phi_{\textmd{top}})}=:S_{\textmd{HM}}. \label{eq:SHM}
\end{align}
Then, the exponent of the tunneling rate is given by
\begin{align}
  \mathcal{B}_{\textmd{HM}}&=S_{\textmd{HM}}-S_{E}\qty[\phi_{\textmd
{FV}}] \nonumber \\
  &=\frac{3}{8\pi}\qty(\frac{1}{V(\phi_{\textmd{top}})}-\frac{1}{V(\phi_{\textmd{FV}})}).
\end{align}
As we wrote above, the difference of the on-shell Euclidean action $\mathcal{B}_{\textmd{HM}}$ governs the magnitude of the transition probability from the false vacuum to the top of the barrier. 
We can see that the transition amplitude for the HM bounce depends only on the values of the potential energy at the false vacuum and the top of the barrier.

In Ref.~\cite{Brown:2007sd}, the authors interpreted this solution as follows. 
If the gap of the potential between the false vacuum and the top of the barrier is negligible compared with their absolute values
\begin{align}
\frac{V(\phi_{\textmd{top}})-V(\phi_{\textmd{FV}})}{V(\phi_{\textmd{FV}})}\ll1,
\end{align}
the transition rate can be approximated with
\begin{align}
  \Gamma_{\textmd{HM}}:=\exp(-\mathcal{B}_{\textmd{HM}}/\hbar)&\simeq \exp\qty(-\frac{\Delta E}{T_{H}}). \label{eq:BHM}
  \end{align}
Here, we have defined the Bekenstein-Hawking temperature of the dS space
\begin{align}
  T_{H}:=\frac{\hbar}{2\pi}H_{s},
\end{align}
and the increment of the vacuum energy
\begin{align}
 \Delta E:=\frac{4\pi}{3}\frac{V(\phi_{\textmd{top}})-V(\phi_{\textmd{FV}})}{H_{s}^3},
\end{align}
where we have used Eq.~\eqref{eq:Hs}.
Eq.~\eqref{eq:BHM} implies that the transition amplitude of the HM bounce is governed by the Boltzmann factor associated with the Bekenstein-Hawking temperature of the dS horizon.
From this perspective, the transition amplitude for the HM bounce can be understood as the probability of finding the homogeneous configuration of the scalar field at the maxima of the potential in the thermal system of $T=T_H$.
In that sense, the HM bounce is often interpreted as a "thermal" transition. Also, as this Boltzmann factor was obtained via the Fokker-Planck equation with the stochastic noise of a homogeneous field in the dS spacetime \cite{Linde:1991sk}, we assume that the stochastic noise can be identified as the thermal fluctuations. On the other hand, to get some insight into the remote HM bounce Ref.~\cite{Weinberg:2006pc,Brown:2007sd}, we will here consider the quantum tunneling effect induced by the quantum fluctuation with superhorizon scale, which is distinct from the thermal (or stochastic) effect, which appears only when interactions between IR and UV modes are turned on \cite{Starobinsky:1994bd}.

Before concluding this subsection, let us mention the applicability condition of the conventional HM amplitude although it was described in the Introduction. As evident from Eq.~\eqref{eq:BHM}, the transition rate depends solely on the values of the Hubble parameter and the potential energy gap and is independent of the field separation between the bottom and top of the potential.
Consequently, this expression suggests that transitions to distant vacua are as likely as those to a neighboring one, provided that those energy gaps are the same.
However, this counter-intuitive {\it remote} HM picture might be invalid, as the overlapping of the wave packet in the false vacuum and that at a distant potential top is vanishingly small, as discussed in Ref.~\cite{Weinberg:2006pc}.
Therefore, we conclude that as was pointed out in Ref.~\cite{Weinberg:2006pc}, the formula~\eqref{eq:BHM} would give the correct transition amplitude for a transition from a false vacuum to a nearby potential top. In Sec.~\ref{Res1}, we come back to this point and provide more detailed discussions there.

\section{Lorentzian formalism}
\label{Lorentzian}

In this section, we apply the Lorentzian path integral to estimate the transition probability for the up-tunneling process, including the remote HM transition.
We shall focus on cases such that the scalar field depends only on the time coordinate $\phi=\phi(t)$.
Also, we consider the quantum-tunneling effect of $\phi(t)$, and the stochastic nature of the super-horizon mode is neglected as we are interested in the remote HM transition or the hopping to the top of the sharp potential barrier, for which the stochastic noise would be suppressed.
Indeed, it is implied in Refs.~\cite{Weinberg:2006pc,Linde:1991sk} that the stochasticity of the fluctuation would not contribute much to such transitions.

Under the assumption of homogeneity, the action can be written as
\begin{align}
\begin{split}
    S[\phi, \rho]&= \int d^4x\sqrt{-g}\qty[\frac{R}{16\pi}-\frac{1}{2}\partial_{\mu}\phi(t)\partial^{\mu}\phi(t)-V(\phi)]\\
    &=2\pi^2\int dt\qty[\rho^3\qty(\frac{1}{2}\dot\phi^2-V(\phi))+\frac{3}{8\pi}(\rho-\rho\dot{\rho}^2)],
\end{split}
\label{lorentzian_act}
\end{align}
where $R=\frac{6}{\rho^2}\qty(\dot{\rho}^2+\rho\ddot{\rho}+\rho)$ is the Ricci scalar and the dot denotes the derivative with respect to $t$.
Introducing the lapse function $N$ by $dt=Nd\lambda$, the action (\ref{lorentzian_act}) reduces to
\begin{align}
    S[\phi, \rho]=2\pi^2\int d\lambda\qty[\rho^3\qty(\frac{1}{2N}\phi'^2-NV(\phi))+\frac{3}{8\pi}\qty(N\rho-\frac{1}{N}\rho\rho'^2)],  
    \label{eq:LS}
\end{align}
where $\lambda$ is the proper time and the prime denotes the derivative with respect to $\lambda$. Throughout this work, we consider the transition amplitude between the initial state at $\lambda=0$ and the final one at $\lambda=1$.

By varying the action with $\phi$, we obtain the EoM for the scalar field
\begin{align}
    \frac{1}{N^2}\qty(\phi''+3\frac{\rho'}{\rho}\phi')=-V_{\phi}, \label{eq:EoM}
    \end{align}
and from the variation with $N$, we obtain the Hamiltonian constraint
\begin{align}
    \frac{1}{N^2}\frac{\rho'^2}{\rho^2}=-\frac{1}{\rho^2}+\frac{8\pi}{3}\qty(\frac{1}{2N^2}\phi'^2+V(\phi)). \label{eq:Hconst}
    \end{align}
Eqs.~\eqref{eq:EoM} and \eqref{eq:Hconst} are nothing but the Lorentzian counterparts of Eqs.~\eqref{eq:phieq} and \eqref{eq:rhoeq}.

\subsubsection{Obtaining classical solution}
\label{Classical}

We here work on the Lorentzian picture and have to obtain a relevant dynamical solution. However, in general, it is hard to solve the equations~\eqref{eq:EoM} and \eqref{eq:Hconst}, and thus we will make two assumptions.
We assume that (i) the potential $V(\phi)$ can be decomposed into $V(\phi) = \Delta V(\phi) + V (\phi_i)$ and $|\Delta V| \ll V (\phi_i)$ holds, where $\phi_i$ is the initial field value of the scalar field. 
Also, it is assumed that (ii) the kinetic term of the scalar field is much smaller than the potential term
\begin{align}
\frac{1}{2N^2}\phi'^2\ll V(\phi). \label{eq:phidV}
 \end{align}
In this situation, the background geometry, relevant to $\rho$, is regarded as the dS background with its Hubble parameter $H_l$:

\begin{align}
   &\rho=\frac{1}{H_{l}}\cosh{(H_{l}N\lambda)}=:\bar{\rho}, \label{eq:HAssume}\\
    &H_{l}:=\sqrt{\frac{8\pi}{3}V(\phi_{i})}. \label{eq:Hl}
 \end{align}
Under the assumptions (i) and (ii), the constraint~\eqref{eq:Hconst} is manifestly satisfied and the EoM for $\phi$ is simplified as
\begin{align}
    \frac{1}{N^2}\phi''+3\frac{H_{l}}{N}\tanh{(H_{l}N\lambda)}\phi'=-V_{\phi},\label{eq:SEoM}
    \end{align}
which corresponds to the Klein-Gordon equation of a homogeneous scalar field in a closed dS spacetime.
Note that we keep the time derivative of the scalar field in contrast to the Euclidean path integral (see  Eq.~\eqref{time_derivative}).
By solving Eq.~\eqref{eq:SEoM} with boundary conditions that are consistent with our assumptions (i) and (ii), we obtain the solution $\bar{\phi}(\lambda)$ which is approximately consistent with the Hamiltonian constraint~\eqref{eq:Hconst}.

\subsubsection{Obtaining on-shell action and steepest descent path}
\label{onshell}

In this part, we discuss how to evaluate the transition rate from the solution.
We allow the scalar field to fluctuate around the solution $\bar{\phi}$ with a small amplitude $\delta\phi$ with conditions $\delta\phi(0)=0=\delta\phi(1)$. It corresponds to the first-order quantum corrections to the semi-classical solution.
Plunging it into the action~\eqref{eq:LS}, we obtain the action at the leading and second-order level as
\begin{align}
    &\bar{S}:=S\qty[\bar{\phi}+\delta\phi, \bar{\rho}]=S_{0}\qty[\bar{\phi}, \bar{\rho}]+S_{2}\qty[\delta\phi, \bar{\rho}], \\
    &S_{0}\qty[\bar{\phi}, \bar{\rho}]=2\pi^2\int_{0}^{1} d\lambda\qty[\bar{\rho}^3\qty(\frac{1}{2N}\bar{\phi}'^2-NV(\bar{\phi}))+\frac{3}{8\pi}\qty(N\bar{\rho}-\frac{1}{N}\bar{\rho}\bar{\rho}'^2)], \\
    &S_{2}\qty[\delta\phi, \bar{\rho}]=2\pi^2\int_{0}^{1} d\lambda\frac{\bar{\rho}^3}{2N}(\delta\phi')^2.
    \end{align}

Thus, in the semi-classical approximation, the propagator for the field transition from $\phi_{i}$ to $\phi_{f}$ is given as
\begin{align}
    G[\phi_{i};\phi_{f}]&=\int_{0}^{\infty}\mathcal{D}N\int_{\phi(0)=\phi_{i}}^{\phi(1)=\phi_{f}}\mathcal{D}\phi\exp\qty(i\frac{S}{\hbar}) \nonumber \\   &\simeq\int_{0}^{\infty}\mathcal{D}Ne^{i\frac{S_{0}}{\hbar}}\int_{\delta\phi(0)=0}^{\delta\phi(1)=0}\mathcal{D}(\delta\phi)e^{i\frac{S_{1}}{\hbar}}  \nonumber \\
    &=\int_{0}^{\infty}\mathcal{D}N A(N)e^{i\frac{S_{0}}{\hbar}},
    \end{align}
where $A(N)$ is a function of $N$ and corresponds to the prefactor of the transition rate in vacuum decay (see e.g. Ref.~\cite{Hayashi:2021kro}).
In order to evaluate the leading contribution to $G[\phi_{i};\phi_{f}]$, we shall deform the contour of $N$ from the real axis to the steepest descent paths following the Picard-Lefschetz theory.
On the new contour, $\mathcal{J}$, which intersects with the relevant steepest descent, the action is dominated by the value at the saddle point $N=N_s$ where $\left.\partial S_{0}/\partial N\right|_{N=N_{s}}=0$.
Thus, the propagator is approximated by
\begin{align}
    G[\phi_{i};\phi_{f}]&\simeq\int_{\mathcal{J}}\mathcal{D}N A(N)e^{i\frac{S_{0}}{\hbar}}  \nonumber \\
    &\simeq\sum_{s}n_{s} A(N_{s})e^{i\frac{S_{\textmd{eff}}}{\hbar}},
    \end{align}
where $s$ is the label of each relevant saddle point of the action in the complex-$N$ plane,  $N_{s}$ is the value of the lapse function at a saddle point, $n_{s}$ is an integer of $0$ or $\pm 1$ depending on the orientation of the contour over the thimbles, and $S_{\textmd{eff}}:=S_{0}(N_{s})$.
From this expression, we can see that the leading contribution of the propagator (the transition rate) is given by $\text{Im}[S_{\textmd{eff}}]$.

To assess the impact of quantum fluctuations around the semi-classical solution on the transition, $A(N_{s})$, numerical integration of $S_{1}$ is required.
However, this poses a technical challenge.
Therefore, for the rest of this paper, our focus will be on evaluating the effective semi-classical action $S_{\textmd{eff}}$, which serves as the counterpart to $\mathcal{B}$ in the Euclidean path integral formalism and is expected to be the primary contribution to the tunneling process.

\section{Example: linear potential}
\label{Linear} 

In the previous section, we introduced the method to evaluate the transition rate of the homogeneous scalar field using the Lorentzian path integral.
In this section, as an example, we perform the Lorentzian path integral for an up-tunneling process in a simple effective potential and obtain the transition amplitude $\exp\qty(-\text{Im}[S_{\rm eff}]/\hbar)$.

\subsection{Setting}
\label{Set}

\begin{figure}[htbp]
    \begin{center}
      \includegraphics[width=8cm]{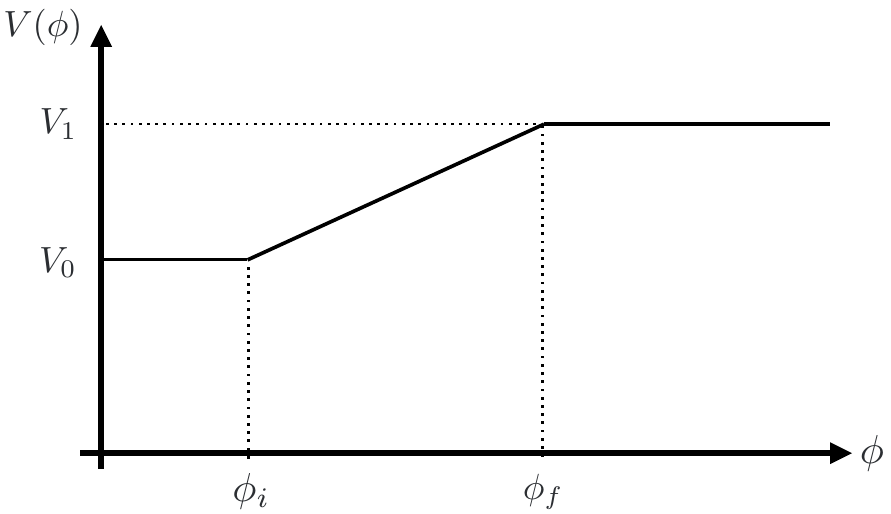}
      \caption{The scalar potential which is constructed by the piecewise of the flat and the linear potentials~\eqref{eq:casesV}.
      The transition we discuss is the quantum jump from $\phi_{i}$ to $\phi_{f}$.}
      \label{fig:linpot}
    \end{center}
  \end{figure}

As we are interested in a homogeneous up-tunneling process in dS spacetime, we can use a potential with a false vacuum and a hiltop. 
Analyses of vacuum decay usually involve a potential that is differentiable anywhere.
However, such potential makes Eq.~\eqref{eq:SEoM} challenging to solve analytically and to obtain the on-shell action.
As such, in this paper, we will explore a simplified and solvable model. 
We here focus on a potential constructed with some linear functions:
\begin{equation}  \label{eq:casesV}
    V(\phi)=
        \begin{cases}
            V_{0}   &   (\text{$\phi\leq\phi_{i}$})  \\
            V_{0}+\frac{\phi-\phi_{i}}{\phi_{f}-\phi_{i}}(V_{1}-V_{0})  &   (\text{$\phi_{i}<\phi\leq\phi_{f}$}) \\
            V_{1}   &   (\text{$\phi_{f}<\phi$})
        \end{cases},
    \end{equation}
which is depicted in Fig.~\ref{fig:linpot}.
In this model, $\phi_{i}$ and $\phi_{f}$ correspond to the false vacuum and the potential top, respectively, and we shall solve the EoM to obtain an analytic solution that describes the upward tunneling from $\phi_{i}$ to $\phi_{f}$ with desired boundary conditions. One can also change the separation $|\phi_f - \phi_i|$ to consider a long field separation while fixing the height to discuss the remote HM transition.

\subsection{Classical action and saddle point}
\label{Res1}

When the potential is given as~\eqref{eq:casesV}, the EoM~\eqref{eq:SEoM} can be solved with the boundary conditions $\phi(\lambda=0)=\phi_{i}$ and $\phi(\lambda=1)=\phi_{f}$, as
    \begin{align}
    \begin{split}
    \phi(\lambda)&=\frac{\tan ^{-1}(\sinh (H_{l}N\lambda )) \qty[-3H_{l}^2(\Delta\phi)^2+\Delta V\log (\cosh (H_{l}N))+\Delta V]}{3H_{l}^2 \Delta\phi\qty[\tan ^{-1}(\sinh (H_{l}N))+\tanh (H_{l}N)
   \text{sech}(H_{l}N)]} \\
   &+\frac{\tanh (H_{l}N
   \lambda ) \text{sech}(H_{l}N \lambda ) \qty[-3 H^2 \qty(\Delta \phi)^2-\Delta V\text{sech}^2(H_{l}N)+\Delta V \log
   (\cosh (H_{l}N))+\Delta V]}{3H_{l}^2 \Delta\phi\qty[\tan ^{-1}(\sinh (H_{l}N))+\tanh (H_{l}N)
   \text{sech}(H_{l}N)]} \\
   &+\frac{\text{sech}(H_{l}N) \qty[3H_{l}^2 \tanh (H_{l}N) \phi_i \Delta\phi-\Delta V \text{sech}(H_{l}N) \tan^{-1}(\sinh
   (H_{l}N\lambda ))]+3H_{l}^2\phi_i\Delta\phi\tan^{-1}(\sinh (H_{l}N))}{3H_{l}^2 \Delta\phi\qty[\tan^{-1}(\sinh (H_{l}N))+\tanh (H_{l}N)
   \text{sech}(H_{l}N)]} \\
   &=:\bar{\phi}\qty(\lambda;\phi_{i}, \Delta\phi,\Delta V,H_{l},N),
   \end{split}
    \end{align}
where we have defined $\Delta\phi:=\phi_{f}-\phi_{i}$ and $\Delta V:=V_{1}-V_{0}$.
By substituting $\bar{\phi}(\lambda)$ into the action~\eqref{eq:LS}, we obtain the on-shell action $S_{0} (N)$.

Once we obtain the on-shell action, we can numerically look for the relevant saddle points $N_{s}$ and compute the effective action $S_{\textmd{eff}}=S_{0}(N_{s})$. Let us introduce a parameter
\begin{align}
    \gamma:=\frac{\Delta V}{H^2(\Delta\phi)^2}, 
    \label{eq:gamma}
    \end{align}
which corresponds to the value of $|V''|/H^2$ at the top of a differentiable potential barrier. 
This parameter was introduced in Ref.~\cite{HenryTye:2008xu} to characterize the effective curvature of the potential barrier in the triangular potential.
If we regard this parameter as the curvature of the potential top, the condition for the absence of the CDL bounce~\eqref{eq:Vcond} can be rewritten as
\begin{align}
\gamma\leq\gamma_{\textmd{cr}}:=\frac{32\pi}{3}\frac{V_{1}}{H_{l}^2}=\frac{4V_{1}}{V_{0}}. \label{eq:gammacr} 
\end{align}
Here, we have used Eq.~\eqref{eq:Hl}.

Before presenting our results, it should be noted how we model the remote HM transition in our setup. According to Ref.~\cite{Weinberg:2006pc}, the HM amplitude can be interpreted as a transition amplitude when we consider the transition from a false vacuum to the neighboring potential top and the potential is sufficiently flat near the top ($\gamma \ll 1$).
Additionally, the previous work such as Ref.~\cite{Linde:1991sk} assumes that the system is in equilibrium, and then interpret the HM transition as a stochastic process.
To ensure the system is thermalized within the relevant time scale, say $t$, the field distance between the false vacuum and the potential top $\Delta \phi$ should not be so long that the field cannot reach the top from the false vacuum within $t$. 
Assuming that the typical relaxation time to achieve equilibrium state, $t_{\rm eq}$, can be estimated by the time scale of the stochastic dissipation process of the field, one could use the following relation \cite{Linde:1982uu,Starobinsky:1982ee,Vilenkin:1982wt}
\begin{equation}
\Delta \phi \sim \sqrt{Ht_{\rm eq}} \times H,
\label{stochastic_relation}
\end{equation}
where $H$ is the Hubble parameter and $\sim Ht_{\rm eq}$ is the step number of the stochastic process.
One can read $t_{\rm eq} \sim (\Delta \phi)^2/H^3$ from (\ref{stochastic_relation}) and find $t_{\rm eq} \propto 1/\gamma$ provided that $\Delta V$ and $H$ are fixed.
As the conventional HM transition cannot be applicable to a situation with $t \ll t_{\rm eq}(\Delta \phi, H)$, there should be a certain threshold value $\gamma = \gamma_{\textmd{th}} \ll 1$ above which (i.e., $\gamma_{\textmd{th}} < \gamma \ll 1$) the field space around the false vacuum and potential top achieves equilibrium state within the relevant time scale.
On the other hand, with a shallower potential ($\gamma < \gamma_{\textmd{th}}$) and a large $\Delta \phi$, thermal equilibrium is unlikely to be realized around the top of the barrier within the relevant time scale.
Thus, if we want to discuss the issue (I) concerning the quantum transition to the remote potential top, the relevant situation is $\gamma < \gamma_{\textmd{th}}$ with $\Delta V$ being fixed \footnote{Recently, the transition rate of the HM transition is reproduced by the stochastic formalism without the assumption of thermal equilibrium~\cite{Miyachi:2023fss}.
In Ref.~\cite{Miyachi:2023fss}, the time scale of the transition was evaluated as $t\sim 1/H\gamma$ up to the factor which depends on the detailed shape of the potential.
This implies that, for $t\ll t_{\rm eq}$ or $\gamma\ll1$, even if we relax the condition, the effect of the long wavelength transition can be important.}.
On the other hand, for issue (II) concerning quantum tunneling with a steeper barrier ($|V''| \gtrsim H^2$), it is reasonable to consider the region $\gamma \gtrsim 1$, assuming the stochastic noise can be neglected compared to quantum diffusion.
In the following, based on the above discussion, we will evaluate the transition amplitudes for the two situations (I) and (II).

\begin{figure}[htbp]
    \begin{center}
     \includegraphics[clip,width=8cm]{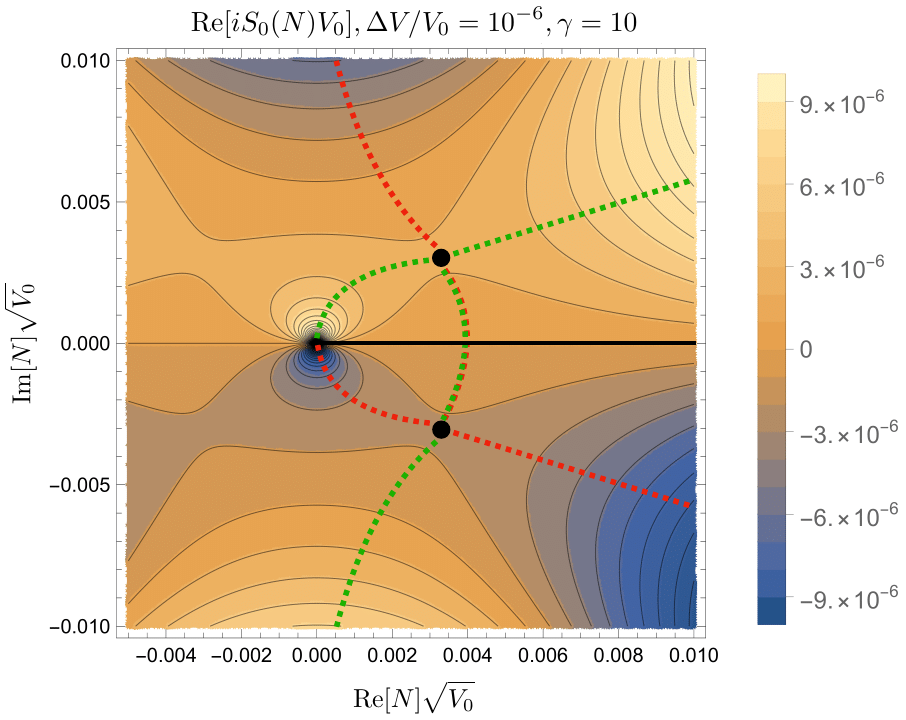} 
     \includegraphics[clip,width=8cm]{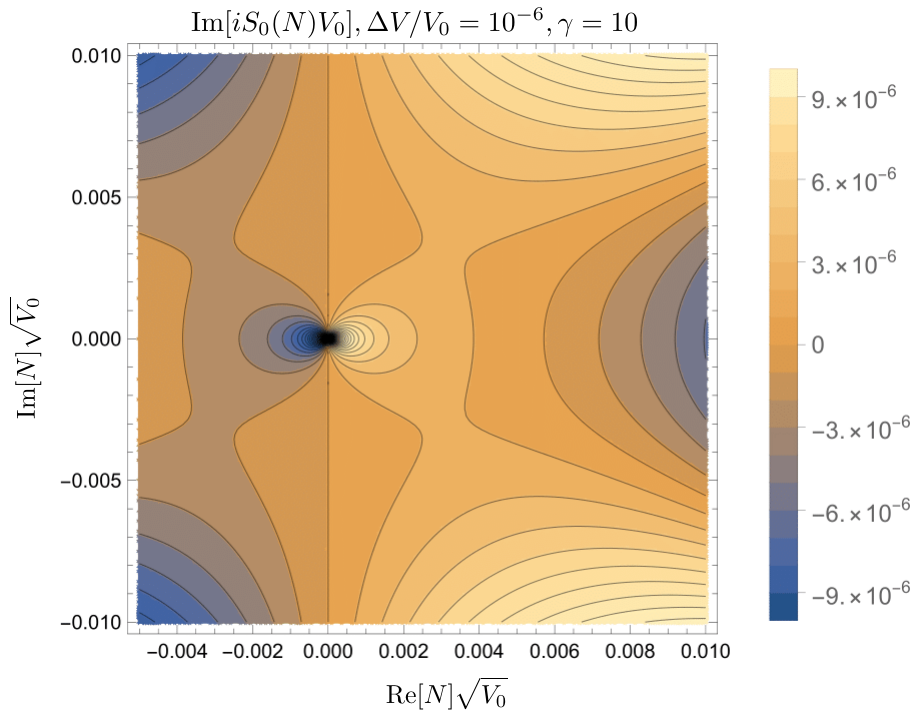} 
     \caption{The contour of $\text{Re}[iS_{0}(N)]$ and $\text{Im}[iS_{0}(N)]$ for $\Delta V/V_{0}=10^{-6}$. $\gamma=10$. 
     In the left figure, the black markers denote the saddle points, and green/red dotted curves represent the steepest ascent/decent paths, respectively.
     The black thick line denotes the original contour $(0, \infty)$.
     }
     \label{fig:dV001}
  \end{center}
   \end{figure}

   \begin{figure}[htbp]
    \begin{center}
     \includegraphics[clip,width=8cm]{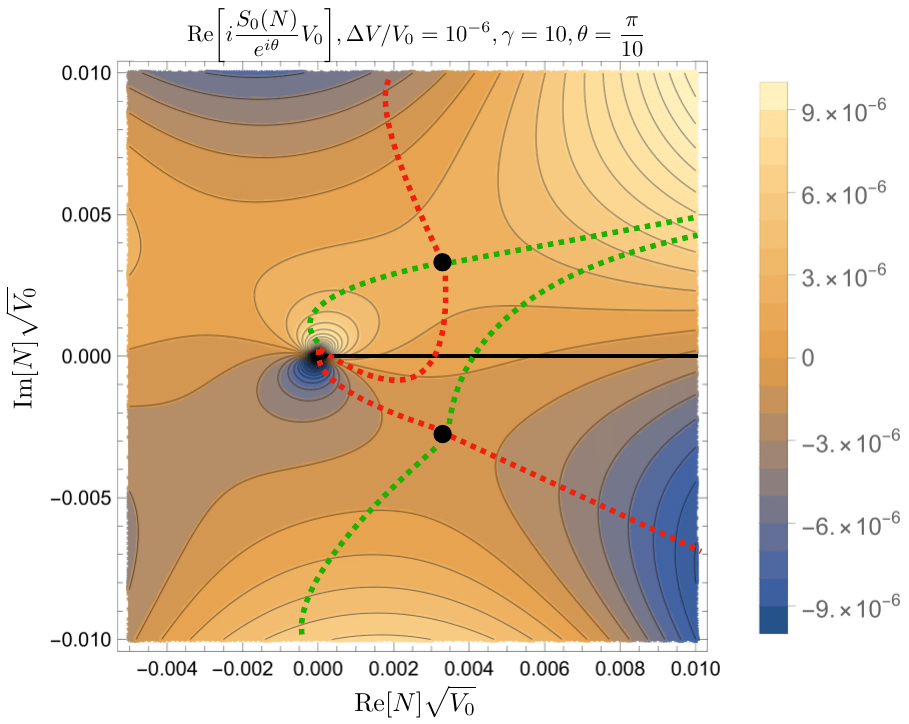} 
     \includegraphics[clip,width=8cm]{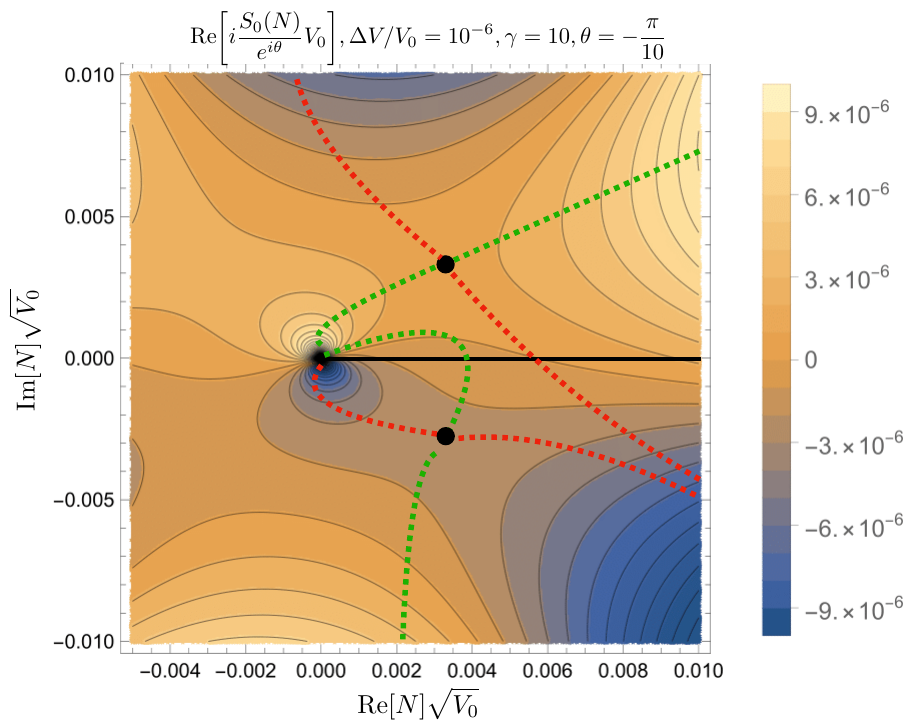} 
     \caption{The contour of $\text{Re}[\frac{i}{\hbar}S_{0}(N)]$ with the substitution $\hbar=e^{i\theta}$.
     The right and left panel denote the cases with $\theta=\frac{\pi}{10}$ and $-\frac{\pi}{10}$, respectively.
     The value of the parameters are fixed same as in Fig.~\ref{fig:dV001}.
     We can see that the degeneracy of the steepest ascent/decent paths are resolved and the steepest ascent path from the lower saddle point crosses with the original contour $(0, \infty)$.}
     \label{fig:dVth}
  \end{center}
   \end{figure}

We present the contour plots of $iS_{0}$ in Fig.~\ref{fig:dV001}.
We can see that saddle points exist in the complex-$N$ plane.
We find two saddle points close to the original contour at $(0, \infty)$ on the real-$N$ axis.
Those saddle points are marked in the left top panel of Fig.~\ref{fig:dV001}.
In the following, we shall denote the saddle points in the upper and lower half region in the complex-$N$ plane as the upper and lower saddle points, respectively.
In the left top panel, the steepest ascent (descent) curves from the saddles are plotted as the green (red) dotted lines.
We observe that one of the steepest descent curves from the upper saddle point coincides with one of the steepest ascent curves from the lower one.
This prevents us from making the one-to-one correspondence between the saddle point and the associated steepest descent and ascent contours, and decomposing the Lefschetz thimbles uniquely.
To resolve this problem, the deformation of the path is useful, as discussed in Refs.~\cite{Feldbrugge:2017kzv,Honda:2024aro}.
Indeed, by taking $\hbar$ to $e^{i\theta}$, as often done in the context resurgence, we can see the degeneracy.
In Fig.~\ref{fig:dVth}, the value of the real part of $\frac{i}{\hbar}S_{0}$ with $\hbar=e^{i\theta}$ is depicted.
We can see that the only one Lefschetz thimble crosses with each saddle point and the steepest ascent curve from the lower saddle point crosses with the original contour $(0, \infty)$, regardless of the signature of $\theta$.
Thus, only the lower saddle point contributes to the dominant transition.

\subsection{Consistency for the slow varying approximation}
\label{Con}

Before evaluating the transition rate, we shall discuss the parameter region consistent with the assumptions made in Sec.~\ref{Classical}.
Regarding assumption (i), we can satisfy it by choosing a value of $\Delta V$, and our focus will be on the parameter range where $\Delta V/V_{0}\geq10^{-2}$.
On the other hand, whether condition (ii) is satisfied or not depends on the on-shell solution, in other words, the result of the numerical computation.
Thus, we shall compare the magnitudes of the kinetic and potential terms derived from the saddle point solution and identify an appropriate parameter range for $\Delta V$.

\begin{figure}[htbp]
    \begin{center}
     \includegraphics[clip,width=16
     cm]{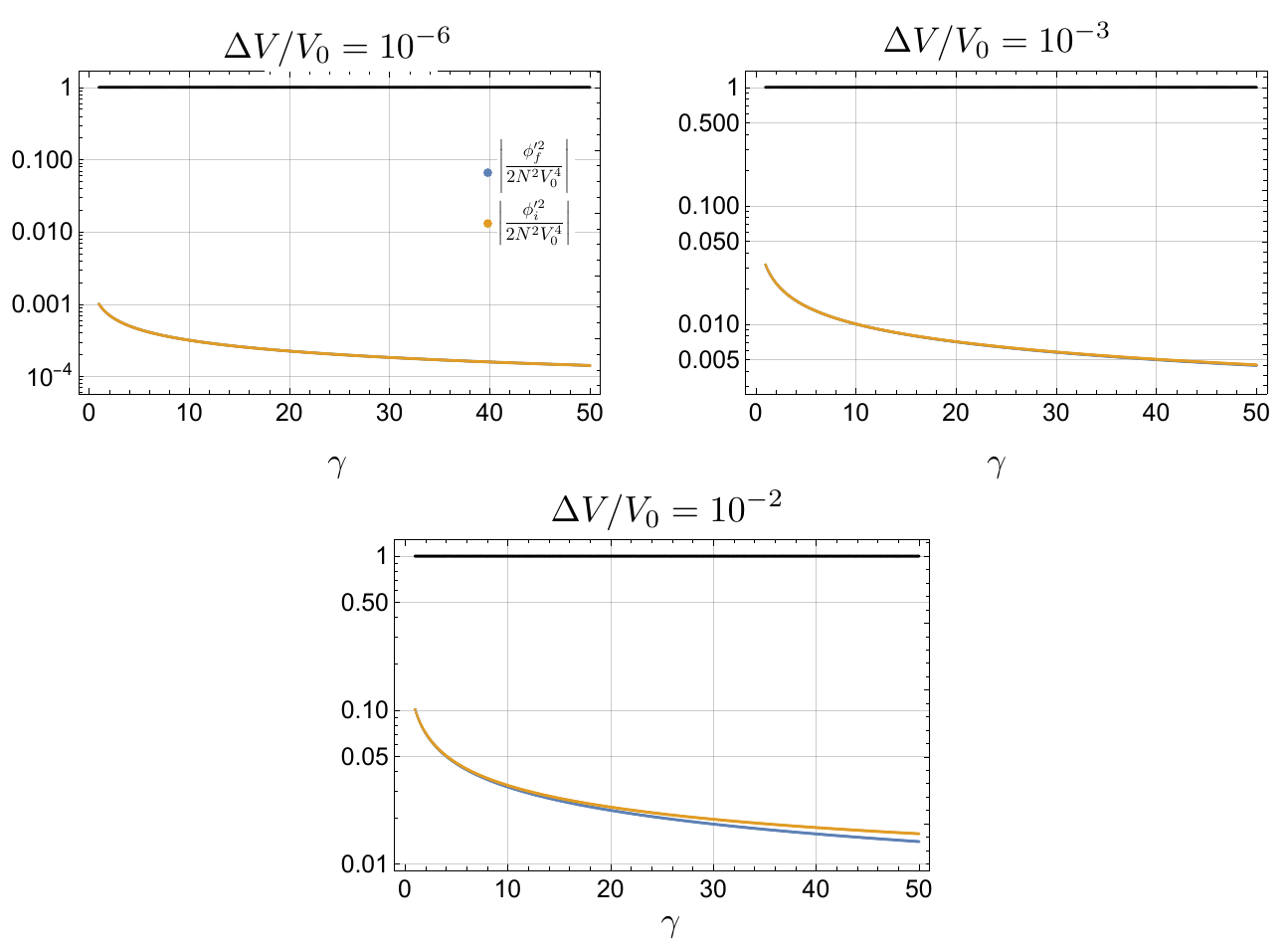} 
     \caption{The ratio of the absolute value of the kinetic term of the field at the initial and final time with the potential energy of the false vacuum, $\left|\frac{1}{2N^2V^4_{0}}\phi'^2\right|$ for $\Delta V/V_{0}=10^{-6}$ (upper left), $10^{-3}$ (upper right) and $10^{-2}$ (lower).
     We have substituted the on-shell solution and the value of the saddle point $N_{s}$ into $\phi'$.
     We also plot the potential energy $V_{0}/V_{0}=1$ (black line).
     We can see that, for $\Delta V/V_{0}=10^{-2}$}
     \label{fig:kin}
  \end{center}
   \end{figure}

In Fig.~\ref{fig:kin}, we illustrate the initial and final kinetic energy $\frac{1}{2N^2}\phi'^2$ with the potential energy $V\simeq V_{0}$.
Here, the on-shell solution $\bar{\phi}$ is substituted into the field velocity $\phi'$.
For the cases of $\Delta V/V=10^{-6}$ and $10^{-3}$ with $\gamma>10$, we observe that the kinetic energy is much smaller than the potential energy, which is consistent with $\frac{1}{2N^2}\phi'^2\ll V(\phi)$.
However, for $\Delta V/V_{0}=10^{-2}$, as depicted in the lower panel in Fig.~\ref{fig:kin}, we find that $\frac{1}{2N^2}\phi'^2\sim O(10^{-1}) V(\phi)$, indicating that the kinetic energy is not negligible compared to the potential energy.
Therefore, for the evaluation with $\gamma=O(1)$, the approximation would not valid in $\Delta V/V_{0}\sim 10^{-2}$.
Considering this result, we restrict our evaluation to the parameter range $\Delta V/V_{0}\leq 10^{-3}$ for consistent evaluation of the transition amplitude.

\subsection{Transition rate}
\label{Res2}

\begin{figure}[htbp]
    \begin{center} 
     \includegraphics[clip,width=18cm]{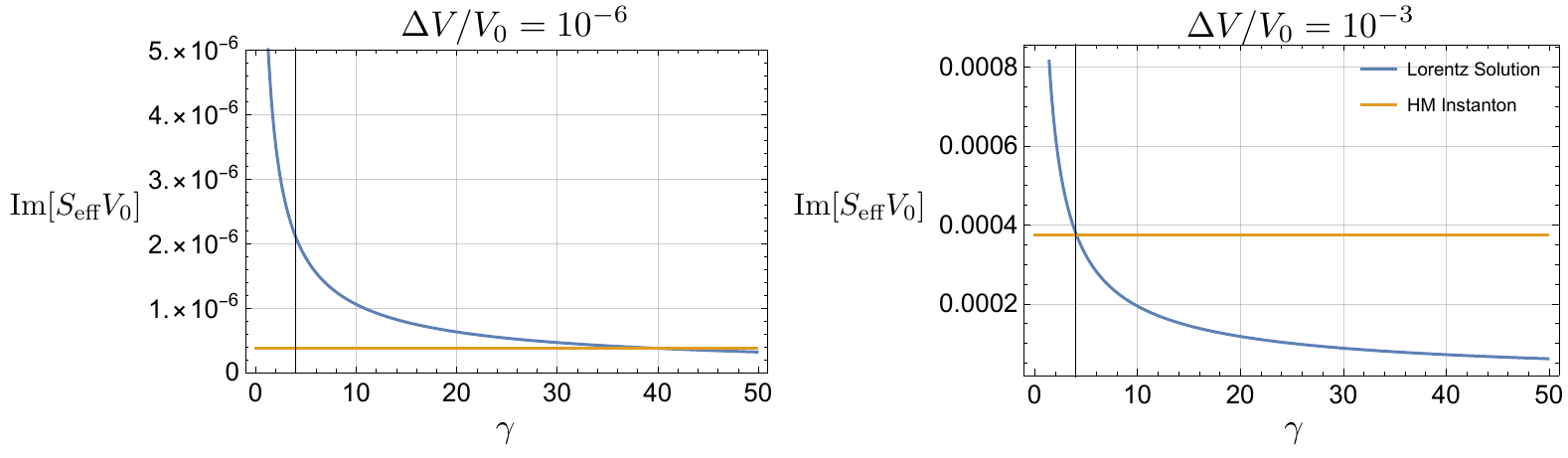} 
     \caption{$\gamma$ dependence of the imaginary part of $S_{\textmd{eff}}$ for $\Delta V/V_{0}=10^{-6}$ (left) and  $10^{-3}$ (right).
     The black solid vertical line is the value of the critical value for $\gamma$, $\gamma_{\textmd{cr}}$.
     We plot the conventional HM transition amplitude (Orange line) and the transition amplitude estimated with the Lorentzian path integral (Blue curve).}
     \label{fig:iSeffdelV}
  \end{center}
   \end{figure}

In this subsection, we examine the $\gamma$ dependence of the transition rate described by the on-shell action. 
As we discussed in the previous subsection, we will restrict our discussion to the parameter range $\Delta V/V_{0}\leq 10^{-3}$, where conditions (i) and (ii) are satisfied.
The values of the imaginary part of $S_{\textmd{eff}}$, which gives the transition amplitude $\exp\qty(-\text{Im}[S_{\rm eff}]/\hbar)$, for $\Delta V/V_{0}\leq 10^{-6}$ and $\Delta V/V_{0}\leq 10^{-3}$ are shown in Fig.~\ref{fig:iSeffdelV}. 
We find that $\text{Im}[S_{\textmd{eff}}]$ depends on both $\Delta V$ and $\gamma$, indicating a non-trivial $\Delta\phi$-dependence of the transition amplitude. 
This is in contrast to the original HM bounce, where the probability is solely characterized by the temperature (Hubble constant) and the energy gap (height of the potential barrier), like the Boltzmann factor in the canonical ensemble. 
This result does not contradict with the original transition amplitude as we do not consider a stochastic process or the free energy in the de Sitter patch, i.e., the on-shell Euclidean action.
We here consider the saddle-point solution on the Lifschetz thimble that does not include stochastic behavior but includes the quantum-mechanical diffusion of the wavefunction, which may be important in the remote Hawking-Moss transition as was discussed in Ref.~\cite{Weinberg:2006pc}.

In Fig.~\ref{fig:iSeffdelV}, we see that for the smaller potential height $\Delta V/V$, the exponent of the transition amplitude  $\text{Im}[S_{\textmd{eff}}]$ is smaller.
Therefore, the transition rate is higher for the smaller energy gap $\Delta V$.
In Fig.~\ref{fig:iSeffdelV}, as references, the exponent of the original HM transition amplitude given by
\begin{align}
   \mathcal{B}_{\textmd{HM}}=\frac{3}{8}\frac{\Delta V}{V_{0}V_{1}},
    \end{align}
is shown with an orange solid line, and the critical value of the curvature, $\gamma_{\textmd{cr}}$, is shown with the vertical black solid line.

In both cases, $\Delta V/V_0 = 10^{-6}$ and $10^{-3}$, shown in Fig.~\ref{fig:iSeffdelV}, the exponent $\text{Im} (S_{\rm eff})$ is equal to or higher than that of the HM bounce for $\gamma<\gamma_{\textmd{cr}}$.
As $\gamma \ll 1$ corresponds to the remote HM transition, it is subtle if the conventional HM amplitude $e^{-{\cal B}_{\rm HM}/\hbar}$ gives the transition amplitude in that limit as discussed in Ref.~\cite{Weinberg:2006pc}.
According to Ref.~\cite{Weinberg:2006pc}, for the remote HM transition, the conventional HM amplitude merely means the probability of finding the homogeneous configuration, but this says nothing about the relative likelihood with the initial configuration.
Thus, when we focus on the homogeneous transition from the false vacuum to the potential top, especially the distant one, the rate may be determined by the quantum tunneling amplitude rather than the Boltzmann factor as the dominant contribution.
If this is the case, the remote HM transition would be suppressed as one can see in Fig.~\ref{fig:iSeffdelV}.
For instance, the exponent of the rate is $\text{Im}[S_{\rm eff}]/{\cal B}_{\rm HM}\simeq89$ for $\gamma=0.1$ and $\Delta V/V_{0}=10^{-6}$. Again, note that the stochastic noise, caused by the interaction between UV and IR modes, is turned off in our computation as we are interested in the short-time transition ($t \ll t_{\rm eq}$) for which the stochastic noise would not play an important role. As such, the deviation from the conventional HM amplitude at $\gamma \sim \gamma_{\rm cr}$ in Fig.~\ref{fig:iSeffdelV} does not imply serious discrepancy or conflict.

Contrary to the result for $\gamma<\gamma_{\textmd{cr}}$, the transition amplitude estimated with the Lorentzian path integral takes a larger value for a steeper potential barrier, $\gamma\geq\gamma_{\textmd{cr}}$, that is another case the conventional HM bounce does not cover.
This is reasonable, at least in the limit of $\gamma \gg \gamma_{\rm cr}$ as the higher value of $\gamma$ means that the slope of the potential barrier is quite thin, or $\Delta \phi$ is shorter, and $\phi$ can reach $\phi = \phi_f$ with a small suppression of the wavefunction. This is consistent with the complementary relation between the HM and CDL bounce solutions, where a steeper potential barrier would suppress the stochastic noises driving the HM transition. Instead, the nucleation of a Coleman (or CDL) bubble can be likely, which is relevant to the quantum diffusion effect or quantum tunneling, not stochastic noise.
Although we cannot make a direct comparison with the Coleman bubble as we here consider a homogeneous scalar configuration, in the case of the nucleation of a true vacuum bubble, the transition amplitude is governed by $e^{- B_{\rm C}/\hbar}$ with~\cite{Coleman:1977py}
\begin{equation}
B_{\rm C} = \frac{27 \pi^2 S^4}{\epsilon^3},
\end{equation}
where $\epsilon$ is the difference between the true and false vacuum energy density, $S \sim \Delta \phi \times \sqrt{V_{\rm top}}$ is the surface tension, and $\sqrt{V_{\rm top}}$ is the height of the barrier. As the tension $S$ is smaller for a thinner potential barrier, the transition amplitude is enhanced for a thinner potential barrier. 

\section{Conclusions and discussions}
\label{Discuss}

Originally, the evaluation of homogeneous tunneling in the de Sitter (dS) background, known as the Hawking-Moss (HM) transition, was performed with the Euclidean path integral methods~\cite{Hawking:1981fz}.
Subsequent works established the derivations of the HM transition amplitude with various schemes (see e.g., Refs.~\cite{Linde:1991sk,Weinberg:2005af,Miyachi:2023fss}). The most common interpretation for the HM bounce is perhaps a thermal excitation induced by stochastic noise.
However, there have been several unclear points regarding the validity and interpretation of this transition picture.
Especially it was pointed out in Ref.~\cite{Weinberg:2006pc} that the conventional HM transition amplitude is not appropriate for the evaluation of the transition between a false vacuum to a distant potential top, which is called a remote HM bounce in Ref.~\cite{Brown:2007sd}.
It was suggested~\cite{Weinberg:2006pc} that in such a case, the HM amplitude may be the Boltzmann factor rather than the transition amplitude, i.e., it just gives the probability of finding the corresponding homogeneous profile in the thermal system with the Bekenstein-Hawking temperature of the dS spacetime.

In this paper, we investigated homogeneous tunneling using the Lorentzian path integral, following the Picard-Lefschetz theory, and evaluated the tunneling probability of a homogeneous scalar field. 
We then discussed the following two issues: (I) 
the contribution of the quantum tunneling effect to the remote HM transition and (II) the HM transition with a steeper barrier for which the thermal effect would be suppressed.
To this end, we turned off thermal/stochastic noises and took into account only the effect of the quantum tunneling with long wavelength by using the Lorentzian path integral.
We have addressed these problems with the following two assumptions: 
(i) that the energy gap between the false vacuum and the potential top is much smaller than the energy density at the false vacuum and (ii) that the kinetic energy of the homogeneous scalar field is much smaller than the energy density at the false vacuum.
These assumptions result in
the closed dS background and
the simplified equation of motion
Eq.~\eqref{eq:SEoM}.

In this paper, we specifically examined a system characterized by the linear potential~\eqref{eq:casesV} as a solvable system.
To characterize the potential structure, we introduced the parameter $\gamma \equiv \Delta V/H^2(\Delta\phi)^2$, which quantifies the curvature around the potential barrier.
When the values of $\Delta V$ and $H$ are fixed, the typical relaxation time for the field to be thermalized, $t_{\rm eq}$, is inversely proportional to $\gamma$ (see the discussion after Eq.~(\ref{stochastic_relation})). 
Therefore, the conventional HM transition amplitude may not be applicable as long as we are interested in up tunneling in the timescale of $t \ll t_{\rm eq}$, which can be regarded as a remote HM transition. Then, we discussed the case (I) relevant to the remote HM transition with $\gamma\ll1$ and fixed $\Delta V$ and $H$. On the other hand, we related the case of $\gamma\geq 1$ to the situation (II) and discussed the HM transition with a steep barrier.
By solving the equation of motion of the scalar field, we obtained the classical on-shell action $S_{0}(N)$ and identified the saddle point, i.e., Lifshitz thimble, on the complex-$N$ plane, where $N$ is the lapse function.
Our analysis determined the one complex Lifshitz thimble $N_{s}$ which contributes to the tunneling process.

We found that the imaginary component of the effective action, $S_{0}(N_{s})$, which determines the order of magnitude of the transition amplitude, monotonically decreases with respect to $\gamma$.
This implies that the transition amplitude for the remote HM transition can be significantly smaller than the conventional HM transition amplitude $\Gamma_{\rm HM}$ when $\gamma \ll 1$.
If this is relevant to the case discussed in Ref.~\cite{Weinberg:2006pc}, the conventional amplitude $\Gamma_{\rm HM}$, which depends only on the initial and final potential energy densities, may not be applicable to the case of $\gamma \ll 1$, as the Euclidean solution merely describes the probability of finding a thermal equilibrium state.
Instead, quantum dispersion causing quantum tunneling may play a dominant role in such remote transitions, although its rate would still be suppressed compared to $\Gamma_{\rm HM}$.
As the original argument was already proposed in Ref.~\cite{Weinberg:2006pc}, we quantitatively and explicitly confirmed this by employing a simpler toy model and the Lorentzian path integral.

Monotonic decreasing of $S_{0}(N_{s})$ also suggests that the higher the curvature around the potential is, the larger the transition probability is.
This implies that the tunneling amplitude is enhanced and overwhelms the thermal transition amplitude for the situation (II), i.e., a sharper and narrower barrier.
This result is reasonable as the barrier becomes thinner for a larger value of $\gamma$, and the quantum tunneling can be completed with the small suppression of the wavefunction.
Although one cannot make a direct comparison with the Coleman de Luccia (CDL) solution as we consider a homogeneous solution here, it is consistent that the false vacuum decay or the CDL decay, relevant to the quantum tunneling process, is more probable than the thermal transition for $\gamma \gg 1$.

In this paper, we only considered one example with the linear potential.
Detailed evaluations, e.g., more complicated potential barrier, for up-tunneling transitions in the Lorentzian path integral need further investigation.
We leave this as a future work.
Also, throughout this paper, we considered homogeneous tunneling in dS spacetime driven by the quantum tunneling effect with long wavelength fluctuation, given that the wavefunction is not decohered even at the superhorizon scale.
However, it is nontrivial whether the quantum nature of the long-wavelength fluctuations is relevant to the dynamics of dS-horizon patches.
This also needs to be investigated by further discussion and computation on the evolution of the primordial fluctuation from other points of view, e.g. quantum information theory.

\section*{Acknowledgements}
D.S. is grateful to T. Miyachi, K. Okabayashi, and J. Tokuda for useful comments and fruitful discussion.
The work of D.S.~was supported by JSPS KAKENHI Grants No.  JP24KJ1223.
The work of N.O.~was supported by Japan Society for the Promotion of Science KAKENHI Grant No. JP23K13111 and by the Hakubi project at Kyoto University.
\bibliography{reference}
\bibliographystyle{unsrt.bst}

\end{document}